# Identity-Based Language Shift Modeling

*Dmitry Ammosov*[1*], *Yalchin Efendiev*[2], *Lenore A. Grenoble*[3]

[1]Chemical and Petroleum Engineering Department, Khalifa University of Science and Technology, Abu Dhabi, 127788, UAE

[2]Department of Mathematics, Texas A&M University, College Station, TX 77843, USA

[3]Department of Linguistics, University of Chicago, Chicago, Illinois, USA

[*]dmitrii.ammosov@ku.ac.ae

**Abstract**

The preservation of endangered languages is a widely discussed issue nowadays. Languages represent essential cultural heritage and can provide valuable botanical, biological, and geographical information. Therefore, it is necessary to develop efficient measures to preserve and revitalize endangered languages. However, the language shift process is complex and requires an interdisciplinary approach, including mathematical modeling techniques. This paper develops a new mathematical model that extends previous works on this topic. We introduce the factor of ethnic identity, which is a proxy for a more complex nexus of variables involved in an individual's self-identity and/or a group's identity. This proxy is socially constructed rather than solely inherited, shaped by community-determined factors, with language both indexing and creating the identity. In our model, we divide speakers into groups depending on with which language they identify themselves with. Moreover, every group includes monolinguals and bilinguals. The proposed model naturally allows us to consider cases of language coexistence and describe a broader class of linguistic situations. For example, the simulation results show that our model can result in cyclic language dynamics, drawing a parallel to cell population models. In this way, the proposed mathematical model can serve as a useful tool for developing efficient measures for language preservation and revitalization.

**Introduction**

The preservation of endangered languages is a highly relevant topic in the modern world. Languages are not only an important cultural heritage [1] but also a source of botanical, biological, and geographical information [2, 3]. Therefore, the development of effective measures for their preservation and revitalization plays an essential role. However, language shift is a complex process that requires careful study with an interdisciplinary approach. Understanding the influence of various factors on the extinction of languages is the key to their preservation and revitalization [4, 5].

Mathematical modeling can be a helpful tool in studying the process of language shift. It allows us to identify the key factors influencing language extinction and to predict language dynamics. There are many works devoted to the development of mathematical models of language shift. However, the model proposed by Abrams and Strogatz is fundamental [6]. The main idea of this model is to determine the probability of transition from one language to another depending on the status of the language and the number of speakers. Subsequent works are mostly devoted to extending this mathematical model to account for various factors such as geography, population growth, and bilingualism [7, 8, 9, 10, 11, 12, 13].

However, the mathematical models noted above predict the disappearance of bilinguals when one of the monolingual populations disappears. Therefore, they predict that only one language survives over the course of a significant time period. At the same time, it is believed that coexistence is possible [14]. To account for such a case, Kandler et al. introduced the "Diglossia model" [15, 16]. In this model, there are additional linear decay terms for each language and linear growth terms for the bilingual population. Such a modification is motivated by social pressure.

In this paper, we propose a new mathematical model that extends previous works and allows us to consider a broader class of language scenarios, including language coexistence. We introduce the factor of ethnic identity, a proxy for more complex variables shaping an individual's self-identity and/or a group's identity [17]. Ethnic identity is socially constructed rather than solely inherited, shaped by community-determined factors, with language both indexing and creating the identity [18, 19]. In our model, we have two groups of speakers identifying themself with one or another language. In each of these groups, there are both monolingual and bilingual populations. Transitions can occur from one language to another (through the bilingual stage) and from one identity to another. The obtained mathematical models can naturally consider language coexistence scenarios. Moreover, our model can describe cyclic language dynamics, which makes a connection with cell population models.

The paper has the following structure. In Section 2, we provide a brief review of previous mathematical models of language shift. Section 3 describes our proposed mathematical model. In Section 4, we present results and discussions. Finally, Section 5 presents conclusions.

**Review of Previous Works**

Mathematical modeling of language shift has gained popularity with the seminal paper by Abrams and Strogatz [6]. In their work, the authors consider two languages (A and B) and denote the population percentages with $a_1$ and $a_2$, where $a_1 + a_2 = 1$. The dynamics are governed by the equations

$$\frac{da_1}{dt} = a_2 \, P_{12}(a_1, s) \, - \, a_1 \, P_{21}(a_2, 1-s),$$

$$\frac{da_2}{dt} = a_1 P_{21}(a_2, 1-s) - a_2 P_{12}(a_1, s). \tag{1}$$

Here, $P_{12}$ refers to the probability that an individual speaking language B converts to language A, and $P_{21}$ is the opposite. The parameter $s \in [0,1]$ refers to the status of the language A. Typical expressions for $P_{12}$ and $P_{21}$ are $P_{12}(a_1, s) \propto a_1^{\beta} s$ and $P_{21}(a_2, s) \propto a_2^{\beta}(1-s)$, $s > 0.5$.

The next generalization of this model includes adding various spatial dynamics, as proposed by Patriarca et al. in [7, 8]. Note Equation (1) includes the dynamics in time only. This includes adding various diffusion and convection terms that disperse the population over a region. The latter is important, but we will not dwell on this. For simplicity, we denote them by $\mathcal{F}_i(a_1, a_2)$.

$$\begin{aligned}\frac{da_1}{dt} &= a_2 P_{12}(a_1, s) - a_1 P_{21}(a_2, 1-s) + \mathcal{F}_1(a_1, a_2), \\ \frac{da_2}{dt} &= a_1 P_{21}(a_2, 1-s) - a_2 P_{12}(a_1, s) + \mathcal{F}_2(a_1, a_2).\end{aligned} \tag{2}$$

The next generalization includes adding population growth, as proposed by Kandler and Steele in [9], modeled by logistic growth functions with certain carrying capabilities (maximum allowed growth). These results are further generalized by varying capacities and making probabilities $P_{ij}$ to be nonlinear functions, as demonstrated by Isern and Fort in [10].

The next set of major generalizations came with the introduction of a model of bilingual population by Mira and Paredes in [11]. The bilingual population is denoted by $a_3$, and the model problem is the following (we factor in spatial dynamics terms in these equations, which were contributed by other authors)

$$\frac{da_1}{dt} = (a_2 + a_3)P_{12}(a_1, s)$$
$$- a_1\left(P_{21}(a_2, 1-s) + P_{31}(a_2, 1-s)\right)$$
$$+ \mathcal{F}_1(a_1, a_2, a_3),$$
$$\frac{da_2}{dt} = (a_1 + a_3)P_{21}(a_2, s) - a_2\left(P_{12}(a_1, s) + P_{32}(a_1, s)\right) \qquad (3)$$
$$+ \mathcal{F}_2(a_1, a_2, a_3),$$
$$\frac{da_3}{dt} = a_1 P_{31}(a_2, 1-s) + a_2 P_{32}(a_1, s) - a_3 P_{12}(a_1, s)$$
$$- a_3 P_{21}(a_2, s) + \mathcal{F}_3(a_1, a_2, a_3).$$

Here, $P_{ij}$ are probabilities going from $j$ to $i$. Note that the authors assume that speakers can switch between languages A to B and also between monolingual to bilingual. Later, Minett and Wang assumed that one can switch between monolingual to bilingual only and modified this model [12].

Later generalizations involve various spatial dynamics terms $\mathcal{F}_i$ for bilingual population, addition of logistic growth models, and introduction of spatially dependent coefficients that can change in some regions (see the work of Castelló et al. in [13]).

These models led to the fact that only one language survives over the course of a significant time period. It is believed that coexistence is possible [20]. Kandler et al. introduced the "Diglossia Model" by adding linear decay terms for each language and, as a compensation, linear growth for the bilingual population (the terms with $w_i$ coefficients) [15, 16]. The latter is argued using social pressure. The model equation has the following form. We repeat this model here, which assumes some simplifications for transition probabilities (we ignore logistic growth terms for simplicity).

$$\frac{da_1}{dt} = -\gamma_{21} a_2 a_1 + \gamma_{13} a_3 a_1 - w_1 a_1 + \mathcal{F}_1(a_1, a_2, a_3),$$
(4)

$$\frac{da_2}{dt} = -\gamma_{12}a_1a_2 + \gamma_{23}a_3a_2 - w_2a_2 + \mathcal{F}_2(a_1, a_2, a_3),$$

$$\frac{da_3}{dt} = (\gamma_{21} + \gamma_{12})a_1a_2 - (\gamma_{13}a_1 + \gamma_{23}a_2)a_3 + w_1a_1 + w_2a_2 + \mathcal{F}_3(a_1, a_2, a_3).$$

These models may lead to the coexistence of mono- and bilingual speakers via the terms in $w$.

**Proposed Mathematical Model**

In our model, we divide the population into two groups. The members of Group 1 ethnically identify with language A, and the members of Group 2 ethnically identify with language B. We introduce the factor of *ethnic identity*, which serves as a proxy for a more complicated nexus of variables involved in an individual's self-identity and/or a group's identity. Ethnic identity is not simply determined by inheritance but is a socially constructed category involving other social variables including race, gender, social class, and status; [17] provides an overview. Which variables are relevant is largely determined by the community or communities involved in the identification. This proxy is used to account for not only inherited (genetic) identity but also ties that can be created by an individual who self identifies with a given group.

There are known ties between language and identity, with language at once used to index identity and also to be a marker of that identity. Language can index not only ethnic identity in the strict sense, but also social class or group membership [21, 22]. Moreover, identity can be created through language [23].

Note that members of each group may speak the other language or be bilingual. Thus, members of Group A may speak Language B and vice versa. We introduce the following notations:

- $a_1$ – the percentage of language A speakers identified as Group 1;

- $a_2$ – the percentage of language A speakers identified as Group 2;
- $b_1$ – the percentage of language B speakers identified as Group 1;
- $b_2$ – the percentage of language B speakers identified as Group 2;
- $c_1$ – the percentage of bilinguals identified as Group 1;
- $c_2$ – the percentage of bilinguals identified as Group 2.

Next, we illustrate the transition between different categories in Fig. 1. For simplicity, we ignore some terms and denote by dashed lines the weak transitions, which will also be ignored. The corresponding equations are (5). This model can easily be enhanced by adding spatial dispersion terms, logistic growth terms, and diglossia terms.

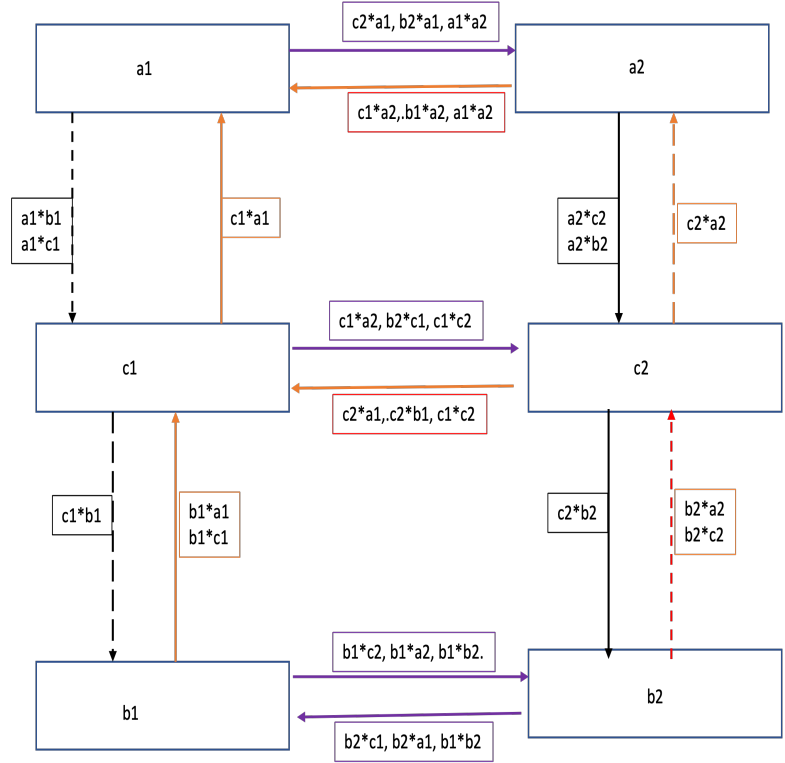

**Fig. 1.** Illustration of main transitions between different categories. The dashed line refers to a weak transition and will be ignored. Some connections depend on several factors that are separated by a comma. The model is simplified

$$\frac{da_1}{dt} = Z_{a1c1}a_1c_1 + Z^a_{a2c1}a_2c_1 + Z^a_{a2b1}a_2b_1 - Z^a_{a1b2}a_1b_2$$
$$- Z^a_{a1c2}a_1c_2 + X_{a1a2}a_1a_2,$$

$$\frac{da_2}{dt} = Z^a_{a1c2}a_1c_2 + Z^a_{a1b2}a_1b_2 - Z^a_{a2c1}a_2c_1 - Z^a_{a2b1}a_2b_1$$
$$- Z_{a2b2}a_2b_2 - Z_{a2c2}a_2c_2 - X_{a1a2}a_1a_2$$
$$\frac{db_1}{dt} = Z^b_{b2c1}b_2c_1 + Z^b_{a1b2}a_1b_2 - Z_{a1b1}a_1b_1 - Z_{b1c1}b_1c_1 \qquad (5)$$
$$- Z^b_{b1c2}b_1c_2 - Z^b_{a2b1}a_2b_1 + X_{b1b2}b_1b_2,$$
$$\frac{db_2}{dt} = Z^b_{b1c2}b_1c_2 + Z^b_{a2b1}a_2b_1 + Z^b_{b2c2}b_2c_2 - Z^b_{b2c1}b_2c_1$$
$$- Z^b_{a1b2}a_1b_2 - X_{b1b2}b_1b_2,$$
$$\frac{dc_1}{dt} = Z_{a1b1}a_1b_1 + Z_{b1c1}b_1c_1 - Z_{a1c1}a_1c_1 + Z^c_{a1c2}a_1c_2$$
$$+ Z^c_{b1c2}b_1c_2 - Z^c_{a2c1}a_2c_1 - Z^c_{b2c1}b_2c_1$$
$$+ X_{c1c2}c_1c_2,$$
$$\frac{dc_2}{dt} = Z_{a2b2}a_2b_2 + Z_{c2a2}c_2a_2 - Z_{b2c2}b_2c_2 - Z^c_{a1c2}a_1c_2$$
$$- Z^c_{b1c2}b_1c_2 + Z^c_{a2c1}a_2c_1 + Z^c_{b2c1}b_2c_1$$
$$- X_{c1c2}c_1c_2.$$

In the next section, we will present numerical results and show that our proposed model can capture various language scenarios.

**Results and Discussion**

In this section, we present some numerical results obtained by simulations of the proposed model (5). Note that a comprehensive numerical analysis of the model is beyond the scope of this paper. We consider the interaction between a bilingual population and a monolingual population of language A (with the disappearance of a monolingual population of language B). As we previously noted, most models assume the disappearance of the bilingual population in such cases.

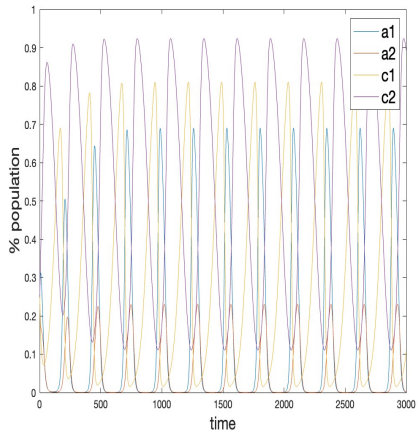 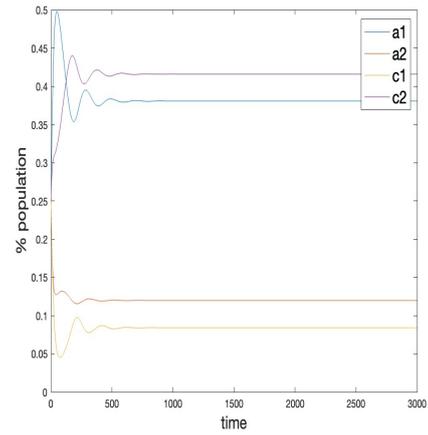

(a)                                              (b)

**Fig. 2.** The dynamics of language A and bilinguals over time. The figure shows cyclic motion

We present the results for different sets of coefficients of the proposed model in Fig. 2 and 3. One can see that our model can predict different scenarios for the coexistence of the bilingual population and the monolingual population of language A. For example, our model can result in cyclic dynamics, i.e., periodically varying populations in each category, as is shown in Fig. 2a, 3a, and 3b. Also, it can predict constant populations in each category (Fig. 2b). Previous approaches can not produce cyclic motions. We note that cyclic dynamics appear in cell population models, which is an indication of cells being alive. In general, it is difficult (mathematically) to understand parameter ranges when cyclic dynamics can occur.

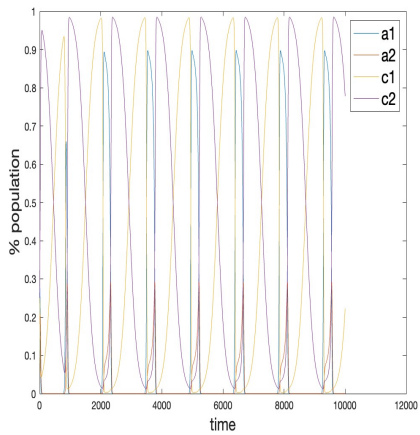 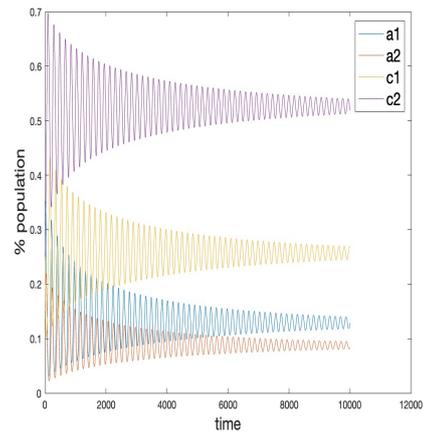

(a)            (b)

**Fig. 3.** The dynamics of language A and bilinguals over time. The figure shows cyclic motion. Different coefficients

**Conclusions**

In this paper, we have proposed a new mathematical model of language shift. We have introduced the ethnic identity factor, a proxy for more complex variables involved in self-identity and group identity. The resulting mathematical model is rich and naturally allows for the coexistence of all languages without adding diglossia terms. Moreover, our equations can result in cyclic dynamics (periodically varying populations in each category), which makes a connection with cell population models. The resulting mathematical model can serve as a useful tool for studying language situations and developing measures to preserve and revitalize endangered languages.

**References**

1. Hale K. Language endangerment and the human value of linguistic diversity. *Language (Baltimore)*. 1992; 68(1):35-42.
2. Harrison K D. *When languages die: The extinction of the world's languages and the erosion of human knowledge*. Oxford University Press; 2008.

3. Nettle D., Romaine S., et al. *Vanishing voices: The extinction of the world's languages*. Oxford University Press on Demand; 2000.
4. Grenoble L., Whaley L. *Saving languages: An introduction to language revitalization*. Cambridge University Press; 2005.
5. Grenoble L. Language contact in the east slavic contact zone. *Balkanistica*. 2015;28:225-250.
6. Abrams D., Strogatz S. Modelling the dynamics of language death. *Nature*. 2003;424(6951):900-900.
7. Patriarca M., Leppänen T. Modeling language competition. *Physica A: Statistical Mechanics and its Applications*. 2004;338(1-2):296-299.
8. Patriarca M., Heinsalu E. Influence of geography on language competition. *Physica A: Statistical Mechanics and its Applications*. 2009;388(23):174-186.
9. Kandler A., Steele J. Ecological models of language competition. *Biological Theory*. 2008;3:164-173.
10. Isern N., Fort J. Language extinction and linguistic fronts. *Journal of the Royal Society Interface*. 2014;11(94):20140028.
11. Mira J., Paredes A. Interlinguistic similarity and language death dynamics. *Europhysics letters*. 2005;69(6):1031.
12. Minett J., SY Wang W. Modelling endangered languages: The effects of bilingualism and social structure. *Lingua.* 2008;118(1):19-45.
13. Castelló X., Loureiro-Porto L., Eguíluz V., Miguel M. S. The fate of bilingualism in a model of language competition. In *Advancing social simulation: the first World Congress.* Springer; 2007. P. 83-94.
14. Edwards J. Bilingualism and multilingualism: Some central concepts. In *The Handbook of Bilingualism and Multilingualism*, Second Edition. Chichester: Wiley-Blackwell; 2012. P. 5-25.

15. Kandler A., Unger R., Steele J. Language shift, bilingualism and the future of britain's celtic languages. *Philosophical Transactions of the Royal Society B: Biological Sciences*. 2010;365(1559):3855-3864.
16. Kandler A., Unger R. Modeling language shift. In *Diffusive spreading in nature, technology and society*. Springer; 2023. P. 365-387.
17. Fought C. Language and ethnicity. In *The Cambridge Handbook of Sociolinguistics*. Cambridge University Press; 2011. P. 238-257.
18. Goffman E. *The Presentation of Self in Everyday Life*. London: Allen Lane; 1969.
19. Waters M. C. The social construction of race and ethnicity: Some examples from demography. *American diversity: A demographic challenge for the twenty-first century;* 2002. P. 25-49.
20. Fishman J. A. Bilingualism and biculturism as individual and as societal phenomena. *Journal of Multilingual and Multicultural Development*. 1980; 1(1): 3–15.
21. Eckert P. Where do ethnolects stop? *International Journal of Bilingualism*. 2008;12(1/2):25-42.
22. Eckert P. Three waves of variation study: The emergence of meaning in the study of sociolinguistic variation. *Annual Review of Anthropology.* 2012;41:87-100.
23. Bucholtz M., Hall K. Language and identity. In Alessandro Duranti, editor, *A Companion to Linguistic Anthropology*. Blackwell, Malden, MA/Oxford; 2003. P. 369-394.